\documentstyle[twoside,fleqn,espcrc2]{article}

\newcommand{\AmS}{{\protect\the\textfont2
  A\kern-.1667em\lower.5ex\hbox{M}\kern-.125emS}}

\title{On a possibility of adjoint colored states condensation at finite
       temperatures in lattice gauge model}

\author{Vladimir K.~Petrov
        \address{Bogolyubov Institute for Theoretical Physics,\\
        National Academy of Sciences of Ukraine, Kiev 143, UKRAINE}}
       
\begin{document}

\begin{abstract}
Cooled down and diluted quark-gluon matter is considered. A possibility of
condensation of multiquark clusters with zero $N$-alities is discussed.
\end{abstract}

\maketitle

\section{Introduction}

Heavy ion experiments are still unable to produce QCD matter that is dense
and hot enough to reveal an explicit evidence in favor of the presence of
quark-gluon plasma. However, a prolonged intermediate period before
hadronisation may give some indirect information about the transient initial
stage.

The common feature of non-Abelian gauge theories is that the static
potential between matter sources crucially depends on the corresponding
representation of the gauge group \cite{dh94}. $q\overline{q}$ states with
zero and nonzero $N$-ality yield screening and confining potentials,
respectively \cite{M77}. In this paper we assume that the retained forces
at intermediate stage rearrange the uniform system and split it into
clusters with zero $N$-ality. Colored states present $\left( N^{2}-2\right)
/(N^{2}-1)$ part of the total number of $q\overline{q}$ states. Although
eventually all multiquark states mandatory rearrange themselves into color
singlets, the requirement that multiquark states must be color singlets may
appear too restrictive at the intermediate stage. Lattice calculations~\cite
{dt85,dt-k87} provide ample evidence that even at fairly high temperatures,
color singlet objects propagate in plasma.  Since the interaction between
droplets with zero $N$-ality is the Debay-like interaction in the confined 
phase and Coulomb-like one in the deconfined phase, one may conclude that 
the gas of such clusters can't be regarded as ideal. Indeed, deviations 
from the ideal gas limit are found even at temperatures of about 
$5T_{c}$ \cite{b-p96}.

Instead of SU(3) we use SU(2) which is expected to have very similar
features. Moreover, many realistic models, e.g. flux tube models, do not
distinguish between SU(2) and SU(3) \cite{Tep:97}.

\section{Gaussian approximation for the effective action}

Let us suppose we succeeded to integrate over spatial link variables $U_{n}$
in the QCD action $S$ and managed to express the effective action $S_{eff}$
in terms of traces of Polyakov loop in fundamental representation 
$\chi _{x}=\mathrm{Tr}_{f}\left\{ \mathrm{\Omega }_{x}\right\} $. 
Then we assume that the Gaussian approximation
\begin{equation}
-S^{eff}\left( \chi _{x}\right) \simeq \eta _{x}\chi _{x}-\frac{1}{2}\chi
_{x^{\prime }}A_{x^{\prime }-x}\chi _{x},  \label{eff2}
\end{equation}
at least roughly, reflects the main features of critical behavior. The
'source' term $\eta _{x}\chi _{x}$ in $\left( \ref{eff2}\right) $ usually
appears (after integration over matter fields) as a part of the effective
fermion action
\begin{equation}
-S_{F}^{eff}\sim 2\sum_{x}\left( \eta \chi _{x}-M^{2}\chi _{x}^{2}\right) .
\label{lin}
\end{equation}

The 'mass' term $M^{2}\chi _{x}^{2}$ in (\ref{lin}) as well
as the invariant measure contribution
\begin{eqnarray}
d\mu _{x} &=& \sqrt{\left( 1-\chi _{x}^{2}/4\right) }\theta 
\left( 4-\chi_{x}^{2}\right) d\chi_{x}/\pi 
\label{3} \\
&\simeq & e^{-\chi_{x}^{2}/8}\theta \left(
4-\chi _{x}^{2}\right) d\chi _{x}/\pi
\nonumber
\end{eqnarray}
are to be included into the matrix $A_{x^{\prime }-x}$. The compactness
condition $\left( \chi_{x}^{2}<4\right)$ can be taken into account in a
spherical model approximation
\begin{eqnarray}
\prod_{x}\theta \left( 4-\chi _{x}^{2}\right)  &\rightarrow &\theta \left(
4v-\sum_{x}\chi _{x}^{2}\right)  \\
&=&\int_{c-i\infty }^{c+i\infty }\frac{ds}{2\pi is}e^{4vs-s\sum_{x}\chi
_{x}^{2}},  \nonumber
\end{eqnarray}
where $v=N_{\sigma }^{3}$ is the spatial lattice volume. This contribution
adds to the $A_{x^{\prime }-x}$ the complementary 'mass' term: $%
-s\sum_{x}\chi _{x}^{2}$, and the integration over $\chi _{x}$ can be easily
done. To obtain the partition function, we integrate over $s$ applying 
the saddle point method
\begin{eqnarray}
2v^{-1}\ln Z\simeq 2s_{0}-\ln \det A\left( s_{0}\right) +\eta _{x^{\prime
}}A_{x^{\prime }-x}^{-1}\left( s_{0}\right) \eta _{x},  
\nonumber
\end{eqnarray}
where $s_{0}$ is the saddle point. Now the matrix $A_{x^{\prime }-x}^{-1}$
can be related to the correlation function
\begin{equation}
A_{x^{\prime }-x}^{-1}=\left\langle \chi _{x}\chi _{x^{\prime
}}\right\rangle -\left\langle \chi \right\rangle ^{2}\
\end{equation}
and expressed through the potential between probes taken from
precision MC simulations
\begin{equation}
V_{1.1}^{\left( f\right) }\left( R\right) /T=-\ln \left\langle \chi _{x}\chi
_{x^{\prime }}\right\rangle =\alpha R-q/R-c ,
\end{equation}
where $\alpha$ is the string tension, and $R=\left| \mathbf{x}-\mathbf{%
x}^{\prime }\right| $. The measured value of $q$ is close to ''IR charge'' $%
q=\pi /12$. In the deconfinement region $\left( \alpha =0\right) $, one can
put $c=-\ln \left\langle \chi _{0}\chi _{\infty }\right\rangle =-\ln
\left\langle \chi \right\rangle ^{2}$.

The potential $V_{1.1}^{\left( A\right) }\left( \left| \mathbf{x}-\mathbf{x}%
^{\prime }\right| \right) $ for $q\overline{q}$ in the adjoint state (for
small $\left| \mathbf{x}-\mathbf{x}^{\prime }\right| $)
\begin{equation}
V_{1.1}^{\left( A\right) }\simeq -T\ln \left( \left\langle {\normalsize \ }%
\chi ^{2}\right\rangle -1\right) 
\end{equation}
becomes to be complex for $\left\langle \ \chi ^{2}\right\rangle <1$ -- 
which means that the adjoint states are strongly suppressed in 
the corresponding parameter area.

Precision data \cite{es99} on $N_{\sigma }^{3}\times 4$ lattices 
($N_{\sigma}=$ 12,18,26,36) show that
\begin{equation}
{\langle }\left| {\chi }\right| {\rangle }=2B_{N_{\sigma }}(\beta /\beta
_{c}-1)^{\varepsilon }  \label{M}
\end{equation}
with $\beta _{c}=2.29895,~~ \varepsilon =0.327$. High statistics calculations
allow us to take away corrections to scaling and to deduce \cite{es99} parameter 
$B_{\infty }\equiv \lim_{N_{\sigma }\rightarrow \infty }B_{N_{\sigma
}}=0.825(1)$ from finite volume data. Considering that (\ref{M}) is in fair 
agreement with entire measured data (up to $\beta =2.3$),
one may assume that $\left( \ref{M}\right) $ gives a reasonable estimation
for $\left\langle \chi ^{2}\right\rangle =\left\langle \left| \chi \right|
\right\rangle ^{2}{+O}\left( 1/v\right) $ in a wider area of $\beta $. In
particular, we find that $\left\langle \chi ^{\left( A\right) }\right\rangle $
becomes positive\footnote{
The potential $V_{1.1}^{\left( A\right) }$ becomes negative for $\beta
>3.7$.} for $\beta >2.8$.

The potential for two ($\mathbf{x}_{1}\simeq \mathbf{x}_{1}^{\prime }\simeq
\mathbf{x}$ and $\mathbf{x}_{2}\simeq \mathbf{x}_{2}^{\prime }\simeq \mathbf{%
x}+\mathbf{R}$) adjoint particles can be computed as
\begin{eqnarray}
\emph{V}^{\left( AA\right) }\left( R\right) &=& F^{\left( AA\right) }\left(
R\right) -F^{\left( AA\right) }\left( \infty \right) 
\label{AA} \\
&=& -2\rho \left( R\right)
q\allowbreak /R,
\nonumber
\end{eqnarray}
where the function $\rho(R)$ slowly changes from $1$ to $
2\left\langle \chi \right\rangle ^{4}/\left\langle \chi ^{\left( A\right)
}\right\rangle ^{2}$; $F^{\left( AA\right) }\left( R\right) \equiv -T\ln
\left\langle \chi _{0}^{\left( A\right) }\chi _{R}^{\left( A\right)
}\right\rangle $ and $F^{\left( AA\right) }\left( \infty \right) =-T\ln
\left\langle \chi ^{\left( A\right) }\right\rangle ^{2}$. Therefore, for any
pair of adjoint particles we get the attractive Coulomb-like potential.

\section{Quasi-ideal gas of adjoint particles}

To obtain the condensation condition, we consider a simple model where
the energy of $n$ adjoint particles is given by
\begin{equation}
E_{n}^{\left( A\right) }=E_{id}^{\left( A\right) }+V_{n}^{\left( A\right)
}\left( x_{1}...x_{n}\right),   \label{fad}
\end{equation}
where $E_{id}^{(A)} = \sum_{k=1}^{n}{\cal E}_{1}\left(p_{k};2m\right)$, 
${\cal E}_{1}(p;m) =$ $\sqrt{p^{2}+m^{2}}-m$ corresponds to 
the kinetic part of energy and $V_{n}^{\left( A\right) }\left(
x_{1}...x_{n}\right) $ corresponds to the potential. Here we make use of the
standard trick and, after the integration over $p_{k}$, write for the 
free energy $F\equiv -T\ln Z$
\begin{equation}
F=F_{id}-T\ln \left\{ 1-v^{-n}\sum_{\left[ x\right] }\left(
1-e^{-V_{n}^{\left( A\right) }/T}\right) \right\}
\end{equation}
with $F_{id}$ $=$ $n$ $\lambda \left( 2m\right)$, where $\lambda \left(
m\right) $ for $m\gg T$ is given by
\begin{equation}
\lambda \left( m\right) =\int e^{-{\cal E}_{1}\left( p;m\right) }\left(
\frac{dp}{2\pi }\right) ^{3}\simeq \left( \frac{mT}{2\pi }\right)^
{\frac{3}{2}}.
\end{equation}

The gas is considered to be so diluted that the scattering of more than two
adjoint particles may be neglected
\begin{equation}
V_{n}^{\left( A\right) }\left( x_{1}...x_{n}\right) \simeq \sum_{jk} 
V^{\left( AA\right) }\left( \left| \mathbf{x_{j}-x}_{k}\right| \right),
\end{equation}
where $V^{\left(AA\right)}\left(R\right) $ is given by (\ref{AA}), so
\begin{eqnarray}
&& \sum_{\left[ x\right] }\left( 1-e^{-V_{n}^{\left( A\right) }/T}\right)
\simeq 
\label{14} \\
&& n\left( n-1\right) \sum_{R>R_{\min }}e^{-2V_{1.1}^{\left( f\right)
}\left( R\right) /T}
\nonumber
\end{eqnarray}
and, therefore, one can write
\begin{equation}
F\simeq F_{id}+n^{2}TB/v;P=\left( 1+Bn/v\right) Tn/v,
\end{equation}
where
\begin{eqnarray}
B &=& -\sum_{R>R_{\min }}^{\sqrt[3]{3v/4\pi }}\left( e^{-V^{\left( AA\right)
}\left( R\right) /T}-1\right) 
\label{16} \\
&\sim & -\frac{3\left\langle \chi \right\rangle
^{4}v^{2/3}}{\left\langle \chi ^{\left( A\right) }\right\rangle ^{2}T}.
\nonumber
\end{eqnarray}
Gas becomes unstable when $\partial P/\partial v\leq 0$, which can be
expressed in terms of the concentration as $n/v\geq Tv^{-2/3}\left\langle
\chi ^{\left( A\right) }\right\rangle ^{2}/\left\langle \chi \right\rangle
^{4},$ so the condensation may start for a very diluted gas .

\section{Area of adjoint states domination}

Let us now try to estimate the value of $\beta $ at which the formation of
adjoint particles begins to dominate. With this in mind we compute the grand
canonical partition functions $Z_{f}\equiv e^{-F^{f}/T}$ and $Z_{A}\equiv
e^{-F^{A}/T}$ for the gas of fundamental and adjoint particles respectively.
The energy for the fundamental particle gas is given by
\begin{equation}
E_{q;\overline{q}}=\left( q+\overline{q}\right) {\cal E}_{1}\left(
m\right) +V_{q;\overline{q}}  \label{qq}
\end{equation}
with $V_{q;\overline{q}}\left( x_{1},...,x_{q},x_{\overline{1}}^{\prime
},...,x_{\overline{q}}^{\prime }\right) $ for the potential energy of 
$q$ quarks and $\overline{q}$ antiquarks. So, we get
\begin{equation}
e^{-F^{f}/T}=\sum_{q,\overline{q}=0}^{\infty }\frac{\lambda \left( m\right)
^{\overline{q}+q}}{q!\overline{q}!}\sum_{\left[ x;x^{\prime }\right]
}e^{-V_{q;\overline{q}}/T}.
\end{equation}
Now after \cite{McLSv}, we may write
\begin{eqnarray}
e^{-\frac{V_{q;\overline{q}}}{T}} &=&\mathrm{Tr}\left\{
e^{-S}\prod_{k}^{q}\chi _{x_{k}}\prod_{\overline{k}}^{\overline{q}}\chi _{x_{%
\overline{k}}^{\prime }}^{\ast }\right\} /\mathrm{Tr}\left\{ e^{-S}\right\}
\nonumber \\
&=&\left\langle \prod_{k}^{q}\chi _{x_{k}}\prod_{\overline{k}}^{\overline{q}%
}\chi _{x_{\overline{k}}^{\prime }}^{\ast }\right\rangle 
\end{eqnarray}
or
\begin{equation}
e^{-F^{f}/T}=\left\langle \exp \left\{ \lambda\sum_{x}\left(
\chi_{x}+\chi_{x}^{\ast}\right)\right\} \right\rangle.
\end{equation}

Along the same line for the adjoin particles, we can get
\begin{equation}
e^{-F^{A}/T}=\left\langle \exp \left\{ \lambda \left( 2m\right) \sum_{x}\chi
_{x}^{\left( A\right) }\right\} \right\rangle.
\end{equation}
To obtain a rough estimation for the parameter area where $F^{A}$ becomes
lower than $F^{f}$, we use the following approximation (instead of the Gaussian
one): $\left\langle e^{Q}\right\rangle \simeq e^{\left\langle Q\right\rangle }
$. If we compare
\begin{eqnarray}
-\frac{F^{A}/v}{\left( 2\pi \right) ^{\frac{3}{2}}\sqrt{T}} &\simeq &\left(
2m\right) ^{\frac{3}{2}}\left( \left\langle \left| \chi \right|
\right\rangle ^{2}-1\right)  \\
-\frac{F^{f}/v}{\left( 2\pi \right) ^{\frac{3}{2}}\sqrt{T}} &\simeq &2\left(
m\right) ^{\frac{3}{2}}\left\langle \left| \chi \right| \right\rangle,
\end{eqnarray}
we conclude that $F^{A}$ $<$ $F^{f}$ in the area where $F^{A}$ becomes
negative, i.e. for $\left\langle \left| \chi \right| \right\rangle >\sqrt{2}$
or  $\beta >3.\,\allowbreak 7$ for SU(2) (see also footnote in Section 2).

We considered a very simple model for cooled quark-gluon matter. It is shown
that at $\beta >2.8$ favorable conditions appear for the creation of Bose
particles with zero $N$-ality. The formation of such clusters dominates at $%
\beta >3.7$. Forces of attraction between such particles facilitate the
condensation which may start even when a gas is very diluted.

\section{Acknowledgment}

I thank Prof. Juergen Engels for providing me with details of his work cited
in \cite{es99}.

\end{document}